\documentclass[12pt,a4paper]{article}
\usepackage{graphicx}
\usepackage{times}
\newcommand{\teff}{\ifmmode T_{\rm eff} \else T$_{\mathrm{eff}}$\fi}
\newcommand{\logg}{\ifmmode \log g \else $\log g$\fi}
\newcommand{\lL}{\ifmmode \log \frac{L}{L_{\odot}} \else $\log \frac{L}{L_{\odot}}$\fi}
\newcommand{\mdot}{$\dot{M}$}
\newcommand{\myr}{M$_{\odot}$ yr$^{-1}$}
\newcommand{\vsini}{$V$ sin$i$}
\newcommand{\vinf}{$v_{\infty}$}
\newcommand{\vturb}{v$_{\rm turb}$}

\newcommand{\vesc}{v$_{esc}$}
\newcommand{\kms}{km s$^{-1}$}
\newcommand{\msun}{\ifmmode M_{\odot} \else M$_{\odot}$\fi}
\newcommand{\mum}{$\mu$m}
\title{\bf UV, optical and near-IR diagnostics of massive stars}
\author{Fabrice Martins$^1$\\
\vspace{1cm}\\
\normalsize $^1$ GRAAL-UMR5024, CNRS \& Universit\'e Montpellier II\\
\normalsize Place Eug\`ene Bataillon, F-34095 Montpellier, France}
\date{\mbox{}}
\begin{document}
\maketitle
\pagestyle{empty}
%
%
\def\bull{\vrule height .9ex width .8ex depth -.1ex}
\makeatletter
\def\ps@plain{\let\@mkboth\gobbletwo
\def\@oddhead{}\def\@oddfoot{\hfil\tiny\bull\quad
``The multi-wavelength view of hot, massive stars''; 39$^{\rm th}$ Li\`ege Int.\ Astroph.\ Coll., 12-16 July 2010 \quad\bull}%
\def\@evenhead{}\let\@evenfoot\@oddfoot}
\makeatother
%
%
\def\beginrefer{\section*{References}%
\begin{quotation}\mbox{}\par}
\def\refer#1\par{{\setlength{\parindent}{-\leftmargin}\indent#1\par}}
\def\endrefer{\end{quotation}}
%
%
{\noindent\small{\bf Abstract:} 
We present an overview of a few spectroscopic diagnostics of massive stars. We explore the following wavelength ranges: UV (1000 to 2000 \AA), optical (4000--7000 \AA) and near--infrared (mainly H and K bands). The diagnostics we highlight are available in O and Wolf--Rayet stars as well as in B supergiants. We focus on the following parameters: effective temperature, gravity, surface abundances, luminosity, mass loss rate, terminal velocity, wind clumping, rotation/macroturbulence and surface magnetic field. 
}
%
%
\section{Introduction}

The development of sophisticated atmosphere codes combined with the regular access to multi--wavelength observational data (from the X--rays to the radio range) allow improved determination of stellar and wind parameters of massive stars. This in turn affects our understanding of these objects. Since massive stars play key roles in different fields of astrophysics (being the progenitors of long--soft GRBs, the producers of most metals heavier than oxygen, important contributors to the release of mechanical energy in the interstellar medium...) it is crucial to be able to accurately constrain their properties. Here, we present a non exhaustive overview of the main spectroscopic diagnostics used to determine the fundamental parameters of massive stars. We restrict ourselves to the UV, optical and near-infrared ranges. The diagnostics we present in the following apply to O and Wolf--Rayet stars as well as B supergiants.

\section{Stellar parameters}

In this section we present the main spectroscopic methods used to determine the stellar parameters: effective temperature, surface gravity, luminosity, surface abundances.

%
\subsection{Effective temperature}
\label{s_teff}

\begin{figure}[ht]
\centering
\includegraphics[height=10cm,width=12cm]{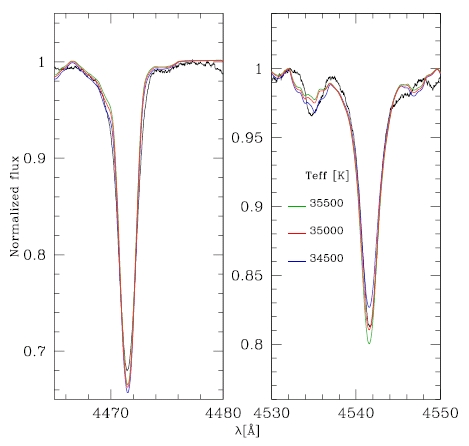}
\caption{Dependence of the HeI 4471 and HeII 4542 line strength on the effective temperature. The synthetic spectra correspond to Teff=34500K (35000K, 35500K) and are shown by the blue (red, green) lines. The observed spectrum is the black line. \label{fig_teff}}
\end{figure}

The effective temperature of massive stars is usually derived using the ionization balance method. The principle relies on the computation of synthetic spectra from atmosphere models at different temperatures. Depending on the temperature, the ionization of the elements present in the atmosphere is different: the wind is more ionized for higher \teff. Consequently, the lines of ions of the same element but of different ionization states are also sensitive to the effective temperature. Comparing the strength of synthetic lines to observed lines thus yields the star's temperature (e.g. Herrero et al.\ 1992, Puls et al.\ 1996, Martins et al.\ 2002). In practice, lines from successive ions of the same elements must be observed. The most reliable indicators for O and Wolf-Rayet stars are the HeI and HeII lines. The classical diagnostics are HeI 4471 and HeII 4542. They are the strongest photospheric lines in most stars (HeII 4686 can be stronger than HeII 4542 but it is more sensitive to wind contamination). An illustration of their behaviour with \teff\ is given in Fig.\ \ref{fig_teff}. We see that increasing \teff\ reduces the HeI 4471 line strength and increases the HeII 4542 absorption. A number of complementary lines can be used to confirm and refine the estimate based on the previously mentioned lines: HeI 4026, HeI 4388, HeI 4712, HeI 4920, HeII 4200, HeII 5412. Note that the HeI singlet lines can be sensitive to subtle details of the modelling  related to line--blanketing effects (e.g. Najarro et al.\ 2006). When the temperature drops below roughly 27000~K, helium is almost neutral in the atmosphere so that no HeII lines are detected. This is the range of mid- and late-B stars. For those objects, one usually switch to the Si ionization balance traced by the following lines: SiII 4124-31, SiIII 4552-67-74, SiIII 5738, SiIV 4089, SiIV 4116 (e.g. Trundle et al.\ 2004). Depending on the temperature, either SiII and SiIII, or SiIII and SiIV lines are used. The temperatures derived from the optical have typical uncertainties of 500 to 2000 K depending on the quality of the observational data and on the temperature itself (uncertainties are larger when lines from one ionization state are weak).

The ionization balance method can be applied to near-IR spectra of O stars. In the K--band, HeI and HeII lines are present above $\sim$ 30000 K (Hanson et al.\ 2005). The strongest HeI lines at 2.058\mum\ must be used very carefully because of its extreme dependence on line-blanketing effects (Najarro et al.\ 1997,2006). The use of HeI 2.112\mum\ is preferred although the line is weaker and can be blended with CIII/NIII emission. The only HeII line in the K-band is HeII 2.189\mum. Repolust et al.\ (2005) have analyzed the same sample of stars independently with optical and near--IR spectra and found that the derived temperatures were consistent within the uncertainties. 

When only UV spectra are available, the determination of \teff\ is more difficult. One usually rely on the iron ionization balance. Line forests from FeIV (resp. FeV, FeVI) are indeed observed in the wavelength range 1600-1630 \AA\ (resp. 1360-1380 \AA, 1260-1290 \AA). An illustration is given in Fig.\ 10 of Heap et al.\ (2006). The {\it relative} strength of these line forests provides the best \teff\ indicator, although the uncertainties are usually larger than those of optical determination.

%
\subsection{Surface gravity}

The surface gravity is classically derived from optical spectroscopy. The wings of the Balmer lines are broadened by collisional processes (linear Stark effect) and are thus stronger in denser atmospheres, i.e. for higher \logg\ (which causes larger pressure and thus more collisions). In practice H$\beta$, H$\gamma$ and H$\delta$ are the main indicators, provided they are in absorption and/or their wings are not contaminated by wind emission. They are usually strong and well resolved.

In the near-IR, the Brackett lines can play the same role. Again, only the wings have to be considered since they are sensitive to collisional broadening. Repolust et al.\ (2005) showed that the behaviour of the Balmer and Brackett lines with gravity was similar only in the far wings, the line cores having different variations (see Repolust et al.\ for a thorough discussion). In practice Br$\gamma$ is the best gravity indicator in the K--band. Br10 and Br11 (H--band) can be used as secondary indicators.

%
\subsection{Luminosity}

Until recently, bolometric luminosities were derived from optical (or near-IR) photometry and bolometric corrections. For instance, one could obtain L$_{\rm bol}$ from $log \frac{L_{\rm bol}}{L_{\odot}} = -0.4 \times (M_{V}+BC(T_{\rm eff})-M_{\odot}^{bol})$ where M$_{V}$ is the absolute magnitude, BC(\teff) the bolometric correction at temperature \teff\ and M$_{\odot}^{bol}$ the sun bolometric magnitude. This method requires the use of calibrations of bolometric corrections. Another related method consists in comparing directly absolute magnitudes (usually in the V band) to theoretical fluxes in the appropriate band convolved with the filter's response.

\begin{figure}[ht]
\centering
\includegraphics[height=9cm]{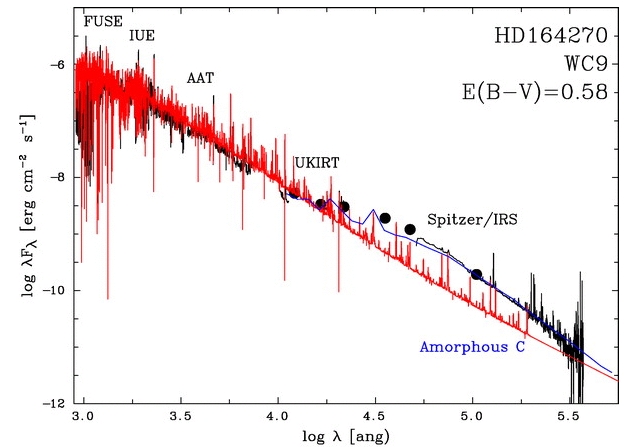}
\caption{SED fitting of a WC9 star. The black line and symbols are the observational data. The red line is the synthetis stellar flux and the blue line is the synthetic dust emission. The infrared part of the spectrum shows an emission excess due to dust. Adapted from Crowther et al.\ (2006a).  \label{figL}}
\end{figure}

Nowadays, sed fitting is becoming the standard way of deriving luminosities. In this process, spectrophotometry ranging from the (far)UV to the infrared is used to adjust the global flux level of atmosphere models. Since the full SED is used, there is no need for bolometric corrections. In addition, the reddening can be derived simultaneously. Any excess emission (due to dust for instance) can be identified and fitted with additional components. An example of such a fit is shown in Fig.\ \ref{figL}. 

For both methods briefly presented above, the distance to the star must be known independently.

%
\subsection{Surface abundances}

Once the effective temperature, gravity and luminosity have been constrained, it is possible to derive the surface abundance of several elements using photospheric lines. The classical spectroscopic method consists in comparing synthetic spectra with different abundances to key diagnostic lines. 

Optical studies of OB stars allow the determination of abundances of C, N, O, Si, Mg. The main diagnostics are the following:

\noindent $\bullet$ carbon: CII 4267, CII 6578--82 / CIII 4647--50, CIII 5696 / CIV 5802--12 \\
$\bullet$ nitrogen: NII 3995 / NIII 4510-15 / NIV 4058, NIV 5200 / NV 4605-20 \\
$\bullet$ oxygen: OII 4075, OII 4132, OII 4661 / OIII 5592 \\
$\bullet$ silicon: SiII 4124--31 / SiIII 4552--67--74, SiIII 5738 / SiIV 4089, SiIV 4116 \\
$\bullet$ magnesium: MgII 4481

\noindent Several NIII lines are also observed between 4630 and 4640 \AA\ in O and early B stars. They are rather strong but their modelling is still difficult and they should be treated with care.

Low ionization lines are present in B stars while high ionization lines are observed in the earliest O stars. The same lines can be used to constrain the abundances of Wolf--Rayet stars. However, they are usually emitted in the wind and are observed in emission. The knowledge of the wind properties, especially the mass loss rate, is thus necessary to correctly derive stellar abundances. 

In O and B stars, the determination of surface abundances requires the knowledge of the microturbulence velocity. It is usually constrained from a few metallic lines, either by direct comparison of synthetic spectra of by measurement of equivalent width of synthetic profiles with different microturbulent velocity. The determination is usually done simultaneously with the abundance determination. The value of \vturb\ is chosen to minimize the spread in abundance derived from several lines of the same ion (e.g. Dufton et al.\ 2005).  

In the UV range, lines from CNO and Si are usually formed in the wind and are used to constrain the mass loss rate and terminal velocity. A determination of the abundances from the optical lines is necessary to correctly derive the wind properties. There are several iron line forests (see also Sect.\ \ref{s_teff}) that can be used to constrain the Fe content. If the relative strength of these line forests constrain \teff, their absolute strength is an indication of the iron composition. 

In the near-IR, the number of metallic lines is limited, especially in OB stars where the lines are weak. For stars with stronger wind (extreme O supergiants and Wolf-Rayet stars) a few features are available. In the K--band, the NIII doublet at 2.247--2.251 \mum\ is a valid indicator of the nitrogen content (Martins et al.\ 2008). The MgII 2.138--2.144 \mum\ doublet is detected in the coolest stars.  In the hottest stars, CIV lines are observed at 2.070, 2.079 and 2.084 \mum. In the H--band, FeII 1.688 \mum, SiII 1.691 \mum\ and SiII 1.698 \mum\ are used to constrain the iron and silicon abundance (Najarro et al.\ 2009).

\section{Wind parameters}

We now turn to the wind parameters of massive stars. We first present the determination of terminal velocities, then the mass loss rates and finally review the spectroscopic diagnostics of clumping.

%
\subsection{Terminal velocity}
\label{s_vinf}

The terminal velocity is the maximum velocity reached by a stellar wind at the top of the atmosphere. If the wind density is high enough, P--Cygni profiles are observed in several lines. The strongest ones are UV resonance lines. The origin of the blueshifted part of the P--Cygni profile is the Doppler shift associated with the wind outflow in front of the photospheric disk. Consequently, the measure of the blueward extent of this absorption gives a direct access to the terminal velocity. The terminal velocity can be defined as the velocity leading to the absorption up to the point where the line profile reaches the continuum (the edge velocity) or as the velocity producing the bluest complete (i.e. zero flux) absorption (the black velocity). The former is usually affected by additional small-scale (microturbulence\footnote{It is introduced as a proxy to represent a significant velocity dispersion which can be simulated by supersonic microturbulence.}) or large--scale (discrete absorption components) motions so that the latter is usually adopted (e.g. Prinja et al.\ 1990). This definition is only valid for strong wind stars though: for thinner winds, the P-Cygni profiles are not saturated. The main UV diagnostics are the following: NV 1240, SiIV, 1393--1403, CIV 1548--50, NIV 1718. Additional indicators are found in the {\it FUSE} range: OVI 1032-1038, CIII 1176. Other P-Cygni profiles can be found below 1000 \AA\ but they are usually blended with interstellar molecular and atomic hydrogen absorption. 

When stars have strong winds (\mdot $\geq 10^{-5}$ \myr) but their UV spectra are not available, other diagnostics can be used. In the optical, the Balmer lines (H$\alpha$, H$\beta$, H$\gamma$, H$\delta$) and sometimes some HeI lines (e.g. HeI 4471) can have pure emission or P-Cygni profiles. In the latter case, the same method as in the UV is applied. For pure emission lines, the line width is usually related to the wind terminal velocity. Fitting such profiles with synthetic spectra computed from atmosphere models with different terminal velocities will provide an indirect measure of the terminal velocity. Similarly, in the near-IR, HeI 2.058\mum\ and HeI 2.112\mum\ have P-Cygni profiles in late-type WR stars or LBVs, and emission profiles in other strong-wind massive stars. We can proceed as for the optical Balmer lines to estimate terminal velocities.

In case no spectroscopic diagnostic is available, the terminal velocity of a massive star can be estimated from the relation \vinf $\approx 2.25 \frac{\alpha}{1-\alpha}$ \vesc where \vesc\ is the escape velocity and $\alpha$ the line force multiplier parameter of the CAK theory. In practice, \vinf $\sim 3 \times$ \vesc\ for stars hotter than about 25000 K, and \vinf $\sim 1.5 \times$ \vesc\ for stars cooler than this limit. This ``bistability jump'' is well known (Lamers et al.\ 1995), although recent studies tend to show that it is more a gradual decrease than a real jump (Crowther et al.\ 2006b, Markova \& Puls 2008).

%
\subsection{Mass loss rate}

There are two main classes of spectroscopic diagnostics of mass loss rate: the P--Cygni resonance lines observed mainly in the UV range, and optical emission lines, mainly H$\alpha$. 

UV P-Cygni profiles are sensitive to the wind density times the ionization fraction of the ion responsible for the observed line. Since the density is directly related to the mass loss rate (density $\propto \frac{\dot{M}}{R^{2}v_{\infty}}$) fitting such features provides constraints on \mdot. The strength of these P--Cygni features allows determinations down to very low values of \mdot\ (typically down to $10^{-10}$ \myr). This is especially important for the so--called 'weak wind stars' (Martins et al.\ 2004, Marcolino et al.\ 2009). The main drawback is that it requires a good knowledge of the ionization structure. All physical processes affecting this structure have to be included in model atmospheres to ensure accurate determinations. The most common features used are: NV 1240, SiIV 1393--1403, CIV 1548--50, HeII 1640, NIV 1718. PV 1118--28 can also be used provided X-rays and clumping are taken into account (see Sect.\ \ref{s_clump}). An example of the fit of the CIV 1548-50 line is shown in Fig.\ \ref{figcl1}. Other lines in the FUV range are available, but they are often contaminated by interstellar atomic and molecular hydrogen absorption (see also Sect.\ \ref{s_vinf}).

\begin{figure}[ht]
\centering
\includegraphics[height=5.5cm]{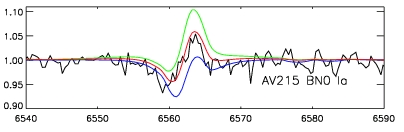}
\caption{H$\alpha$ fit for the SMC B supergiant AV215. The blue and green lines correspond to models with a mass loss rate changed by $\pm$15\% compared to the best fit model (red line). Adapted from Trundle et al.\ (2004). \label{figha}}
\end{figure}

The other main diagnostics of mass loss rate is the H$\alpha$ line in the optical (e.g. Puls et al.\ 1996). If the density is high enough, hydrogen recombination leads to H$\alpha$ wind emission which adds to the underlying photospheric absorption. For strong winds, the emission completely dominates the line profile. Fig.\ \ref{figha} shows an example of fit for a SMC B supergiant (Trundle et al.\ 2004). Since it is a recombination line, it depends on the density square (as opposed to density for P--Cygni profiles). Consequently, its emission decreases quickly with density (and thus \mdot). H$\alpha$ then turns rapidly into a pure photospheric absorption profile from which no \mdot\ determination is possible. This happens below $\sim 10^{-8}$ \myr. H$\alpha$ is however less sensitive to ionization issues since hydrogen is almost completely ionized in massive stars atmospheres. It is thus sometimes considered a better diagnostics (but again, only for strong wind stars). A secondary optical indicator is HeII 4686. In the case of Wolf--Rayet stars, the other Balmer lines (H$\beta$, H$\gamma$, H$\delta$) are also in emission and are complementary indicators.  

In the near-IR range, the Brackett lines, especially Br$\gamma$, play the same role as the Balmer lines in the optical (Repolust et al.\ 2005, Martins et al.\ 2008). A rather strong line is Br$\alpha$ at 4.051 \mum. Preliminary example of use of this line are shown in Puls et al.\ 2008.

%
\subsection{Clumping}
\label{s_clump}

Several pieces of evidence indicate that the winds of massive stars are not homogeneous. Spectroscopically, the first indirect proof came from Hillier (1991) who realized that the red electron scattering wing of strong emission lines of Wolf-Rayet stars was overpredicted in homogeneous models. The inclusion of inhomogeneous winds by means of a volume filling factor approach lead to a better agreement with observations. The electron scattering wings of emission lines are still used nowadays to constrain the degree of inhomogeneities in strong wind stars. The classical diagnostics are: HeII 4686, HeII 5412, H$\beta$ (see Hillier 1991, Martins et al.\ 2009).

The presence of clumping in massive stars winds has two main effects: first, for a given atmospheric structure, it changes the shape of wind lines; second, due to the increased density in clumps, recombinations are stronger and thus the ionization structure is modified. The first effect can be explained as follows. For a recombination line, the line intensity is proportional to $\rho^{2} \times V$ where $\rho$ is the density and $V$ the total volume of the wind. In the case of a volume filling factor $f$, the density in the clumped wind is $\rho_{c}=\rho_{0}/f$ where the indexes 'c' and '0' refer to the clumped and unclumped models respectively. Similarly, the volume effectively containing the material is $f \times V_{0}$. Hence, the line intensity is proportional to $\rho_{0}^{2} / f$. Consequently, including clumping increases the line strength by $1/f$. Said differently, since $\rho_{0} \propto \dot{M}$, the line intensity will be the same for similar $\dot{M}/\sqrt{f}$ ratios. H$\alpha$ in the optical (e.g. Puls et al.\ 2006) and Br10/Br11 in the near--IR (Najarro et al.\ 2009) are the main $\rho^{2}$ diagnostics of clumping.

\begin{figure}[ht]
\begin{minipage}{8cm}
\centering
\includegraphics[height=5.5cm]{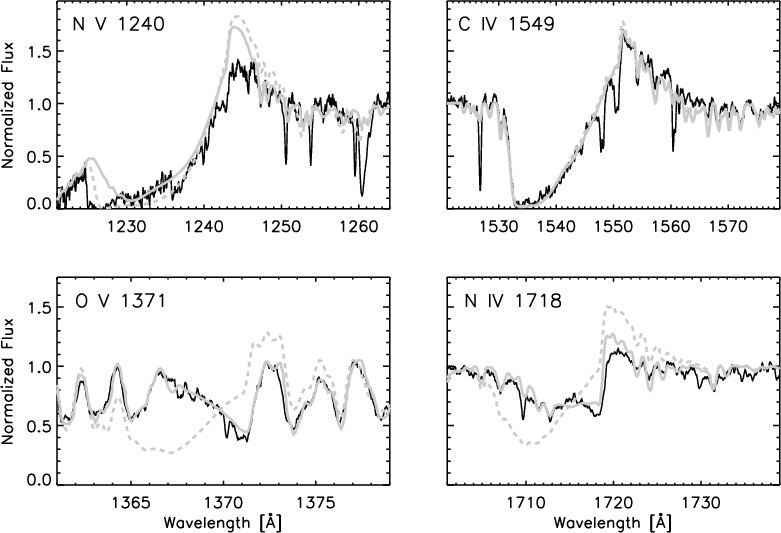}
\caption{Observed P-Cygni profiles (black solid line) together with synthetic spectra from models with homogeneous wind (grey dashed line) and clumped wind (grey solid line). From Bouret et al.\ (2005). \label{figcl1}}
\end{minipage}
\hfill
\begin{minipage}{8cm}
\centering
\includegraphics[height=5.5cm]{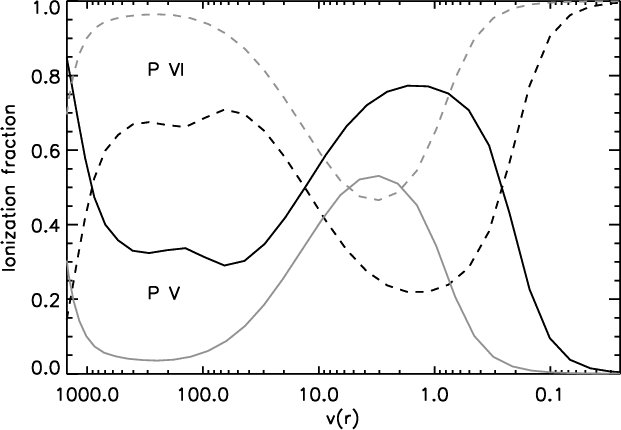}
\caption{Oxygen ionization fraction of the homogeneous (grey) and clumped (black) models shown in Fig.\ \ref{figcl1}. From Bouret at al.\ (2005). \label{figcl2}}
\end{minipage}
\end{figure}

For scattering lines such as the UV P--Cygni profiles, the intensity depends linearly on the density, so that in principle there is no 'first effect' of clumping on these profiles\footnote{This is only true if clumps are optically thin in UV resonance lines (see e.g. Sundqvist et al.\ 2010.)}. But the second effect -- the change of ionization structure -- is present. This is illustrated in Fig.\ \ref{figcl1} and \ref{figcl2} -- from Bouret et al.\ (2005). In the former figure, UV P--Cygni profiles of an O4V((f)) star are shown for homogeneous (grey dashed line) and clumped (grey solid) models. The clumped models provide a much better fit to OV 1371 and NIV 1718. In Fig.\ \ref{figcl2}, we see that adding clumping strongly reduces the fraction of OV in the atmosphere, leading to weaker OV 1371 line. Another UV diagnostic of clumping is the PV doublet at 1118--1128 \AA\ (e.g. Fullerton et al.\ 2006).

More direct evidence for clumping come from time series analysis of selected emission lines of O supergiants and Wolf-Rayet stars. The first study of Eversberg et al.\ (1998) showed the presence of emission sub--peaks on-top of the main emission of HeII 4686. This structures showed motions from the line center to the line wings. This is interpreted as the presence of clumps moving outward in the stellar atmosphere. Similar conclusions were subsequently reached for different types of emission line stars, using CIII 5696 and CIV 5802--12 in addition to the HeII lines mentioned above (e.g. L\'epine et al.\ 2000).

\section{Rotation and magnetic field}

We finally focus on two properties of massive stars: their rotation rates and the relation to macroturbulence, and their magnetic fields.

%
\subsection{Projected rotational velocities and macroturbulence}

The determination of projected rotational velocities (\vsini) has become a difficult task since it was realized that line profiles of O stars were also broadened by another mechanism dubbed macroturbulence. Its origin is not well constrained although a recent study by Aerts et al.\ (2009a) points to the a probable role of stellar pulsations (see also Sim\'on D\'iaz et al.\ 2010 for a first observational evidence).

\begin{figure}[ht]
\centering
\includegraphics[height=9cm]{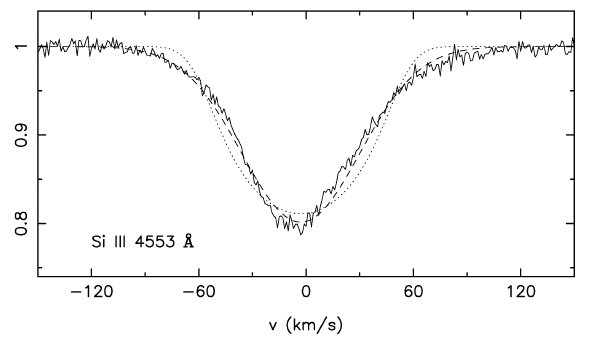}
\caption{Observed profile (solid line) together with a synthetic spectrum including only rotational broadening (\vsini\ = 57 \kms, dotted line) and rotational broadening + Gaussian macroturbulence (\vsini\ = 5 \kms + v$_{\rm mac}$ = 32 \kms, dashed line). The inclusion of macroturbulence leads to a much better fit of the observed profile. From Aerts et al.\ (2009b). \label{figmac}}
\end{figure}

In absence of macroturbulence, two methods have been widely used to constrain \vsini:

\noindent $\bullet$ FWHM--\vsini: this method first developed by Slettebak et al.\ (1975) relies on the computation of synthetic line profiles at different rotational velocities from which the full width at half maximum (FWHM) is measured and subsequently compared to observational data. It was used by Herrero et al.\ (1992) and Abt et al.\ (2002) (among others) to derive \vsini\ for O and B stars. It relies mainly on optical metallic lines.

\noindent $\bullet$ Cross-correlation: here, a low \vsini\ template spectrum is convolved at different rotational velocities and is subsequently cross-correlated with observed spectra. The method has been particularly used in the UV range (e.g. Penny et al.\ 1996, Howarth et al.\ 2007) taking advantage of the large IUE database. 

The direct comparison of synthetic line profiles to observational data revealed that the wings of photospheric lines did not show the classical ``curved'' shape of rotational profiles, but were wider and more ``triangular''. This is illustrated in Fig.\ \ref{figmac} where we see that a pure rotational profile (dotted line) is a poor fit of the observed spectrum (see also Ryans et al.\ 2002). The addition of a macroturbulent profile, usually implemented by convolution with a Gaussian profile and thus mimicking isotropic turbulence, leads to a significant improvement. The consequence is a reduction of the derived \vsini\ compared to studies ignoring macroturbulence (see Fig.\ \ref{figmac}). In practice, optical lines are well suited to constrain \vsini\ and the amount of macroturbulence. Among the key lines, there is: CIV 5812, OIII 5592, NIV 4057, HeI 4712 (see Howarth et al.\ 2007, Martins et al.\ 2010). The main drawback of this method is that several combinations of \vsini/macroturbulence can give fits of similar quality, rendering the determination of projected rotational velocities uncertain.

\begin{figure}[ht]
\centering
\includegraphics[height=10.5cm]{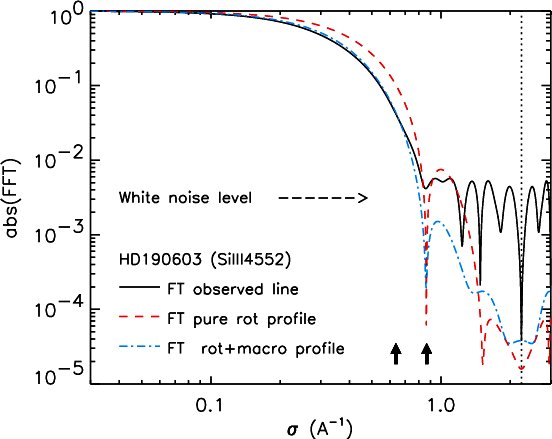}
\caption{Fourier transform of observed and synthetic line profiles of SiIII 4552. The synthetic profiles have the same projected rotational velocity (\vsini\ = 45 \kms) but a different amount of Gaussian macroturbulence (0 for the dashed model, 50 \kms\ for the dot--dashed model). The position of the first zero is the same, independently of the amount of macroturbulence. From Sim\'on D\'iaz et al.\ (2007). \label{figft}}
\end{figure}

A powerful method to break this degeneracy is the use of the Fourier transform (FT) of observed profiles. Provided the macroturbulence is well represented by a symmetric kernel (such as a Gaussian profile), the first zero of the FT is thus directly related to the projected rotational velocity by the relation: $\frac{\lambda}{c} v_{\rm sini} \sigma_{1} = 0.66$ where $\lambda$ is the wavelength of the line center and $\sigma_{1}$ the position of the first zero. An illustration is given in Fig.\ \ref{figft} where one can see that for a given \vsini, the position of the first zero is always the same, regardless of the amount of Gaussian macroturbulence included. Here again, optical metallic lines are well suited for this method: OII 4414, OII 4661, OIII 5592, SiIII 4553, SiIV 4089, NIV 4057, CIV 5812 (e.g. Sim\'on D\'iaz et al.\ 2006). We stress that the conclusion about the relevance of the FT method to derive \vsini\ relies on the assumption that macroturbulence was represented by a symmetric function. If it is not the case (as for pulsations where macroturbulence results from the collective effects of hundreds of oscillations -- see Aerts et al.\ 2009a), then the position of the first zero is affected. The recent study of Sim\'on D\'iaz et al.\ (2010) favour a Gaussian radial--tangential macroturbulence profile over an isotropic Gaussian shape. More analysis are needed to characterize the origin and properties of macroturbulence in massive stars.

%
\subsection{Surface magnetic field}

\begin{figure}[ht]
\centering
\includegraphics[height=11cm,angle=0]{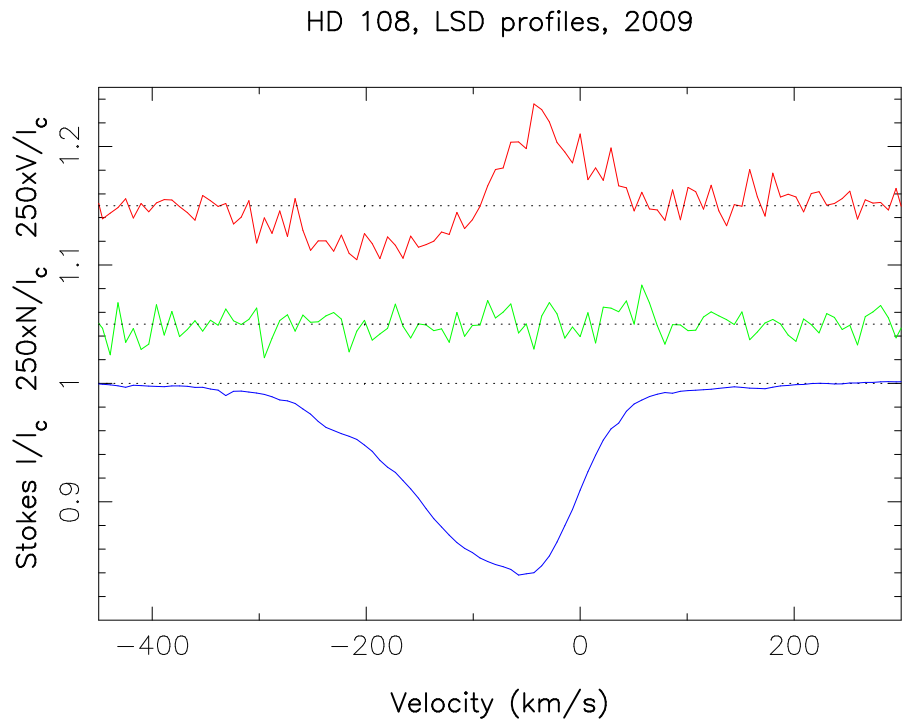}
\caption{Stokes V, N and I profile (from top to bottom) for the Of?p star HD108. The presence of a clear signature in the Stokes V profile and the absence of feature in the null profile (middle) is a direct indication of the presence of the Zeeman effect and thus of a surface magnetic field. From Martins et al.\ (2010). \label{figB}}
\end{figure}

The development of spectropolarimeters working in the optical range has lead to the detection of surface magnetic fields in several O and B stars (e.g. Donati et al.\ 2002, 2006; Bouret et al.\ 2008, Grunhut et al.\ 2009). The principle of the detection relies mainly on the 'least square deconvolution' method (Donati et al.\ 1997). In practice, the idea is to detect Zeeman splitting in photospheric lines. Given the faintness of the polarized signal, a line mask made of several well understood lines is built and an average line profile is created from it (leading to the Stokes I parameter, see Fig.\ \ref{figB}). The detection of a magnetic field is made from the Stokes V profile which is the difference between the right and left circular polarization signal created from the line mask. An example of unambiguous detection is displayed in Fig.\ \ref{figB}. The photospheric lines used to build the line mask are usually the following: HeI 4026, HeI 4388, HeI 4471, HeI 4712, HeII 4200, HeII 4542, NIII 4510, OIII, 5592, CIV 5812. 
Currently, there are no spectropolarimeter working in the infrared range nor in the UV.

\section{Summary}

We have reviewed some of the main spectroscopic diagnostics of massive stars in the UV (1000--2000 \AA), optical (4000-7000 \AA) and near-infrared (H and K bands) wavelength ranges. The description was not exhaustive and was meant to give an overview of the most commonly used spectral lines and methods to derive the stellar and wind parameters of OB and Wolf--Rayet stars. A summary of the main diagnostics is given in Table \ref{tab_sum}. 

\begin{table}
\caption{Summary of the main diagnostic lines for several stellar and wind parameters
\label{tab_sum}} 
\small
\begin{center} 
\begin{tabular}{| l || c | c | c |}
\hline 
              &      UV                     &        Optical                             &       near-IR     \\
\hline
\hline
\teff         &     FeIV/V/VI               &  HeI 4471 / HeII 4542                      &  HeI 2.112 / HeII 2.189 \\
              &                             &  SiII 4124 / SiIII 4552 / SiIV 4116        &                 \\
\hline
\logg         &     --                      &  H$\beta$, H$\gamma$, H$\delta$            &  Br$\gamma$ \\
\hline
\vinf         &   NV 1240, SiIV 1393-1403   &  H$\alpha$, H$\beta$, H$\gamma$, HeI 4471  &  HeI 2.058, HeI 2.112, Br$\gamma$ \\
              &   CIV 1548-1550, NIV 1718   &     {\it (if strong wind)}                 &  {\it (if strong wind)} \\
\hline
\mdot         &   NV 1240, SiIV 1393-1403   &  H$\alpha$, HeII 4686                      &  Br$\gamma$ \\
              &   CIV 1548-50, NIV 1718     &                                            &             \\
\hline
$f$ (clumping)&   OV 1371, NIV 1718         &  H$\alpha$, HeII 4686                      &   Br10, Br11  \\
              &   PV 1118-1128              &                                            &         \\
\hline
Surface       &   FeIV/V/VI                 & CIII 4637-40, CIV 5812,                    & NIII 2.247-2.251 \\
abundances    &                             & NIII 4510-15, NIV 5200,                    & MgII 2.138-2.144 \\
              &                             & OII 4661, OIII 5592...                     & SiII 1.691-98 \\
              &                             &                                            & FeII 1.688, FeII 2.089 \\
\hline
Magnetic      &     --                      & HeI 4026, HeI 4712                         &   --           \\
field         &                             & HeII 4200, HeII 4542,                      &              \\
              &                             & OIII 5592, CIV 5812                        &              \\
\hline 
\end{tabular} 
\end{center} 
\end{table}

%
%
\section*{Acknowledgements}
FM acknowledges financial support from the french P\^ole National de Physique Stellaire (CNRS/INSU). Comments by J. Puls helped to improve this contribution.
%
%
\footnotesize
\beginrefer

\refer Abt H.A., Levato H., Grosso M., 2002, ApJ, 573, 359

\refer Aerts C., Puls J., Godart M., Dupret M.-A., 2009a, A\&A, 508, 409

\refer Aerts C., Puls J., Godart M., Dupret M.-A., 2009b, CoAst, 158, 66

\refer Bouret J.-C., Lanz T., Hillier D.J., Heap, S.R., Hubeny I., Lennon D.J., Smith L.J., Evans C.J., 2003, ApJ, 595, 1182

\refer Bouret J.-C., Lanz T., Hillier D.J., 2005, A\&A, 438, 301

\refer Bouret J.-C., Donati J.-F., Martins F., Escolano C., Marcolino W., Lanz T., Howarth I.D., 2008, MNRAS, 389, 75

\refer Crowther P.A., Morris P. W., Smith J. D., 2006a, ApJ, 636, 1033

\refer Crowther P.A., Lennon D.J., Walborn N.R., 2006b, A\&A, 446, 279

\refer Donati J.-F., Semel M., Carter B.D., Rees D.E., Collier Cameron A., 1997, MNRAS, 291, 658

\refer Donati J.-F., Babel J., Harries T.J., Howarth I.D., Petit P., Semel M., 2002, MNRAS, 333, 55

\refer Donati J.-F., Howarth I. D., Bouret J.-C., Petit P., Catala C., Landstreet J., 2006, MNRAS, 365, L6

\refer Dufton P.L., Ryans R.S.I., Trundle C., Lennon D.J., Hubeny I., Lanz T., Allende Prieto C., 2005, A\&A, 434, 1125

\refer Eversberg T., Lepine S. Moffat A.F.J., 1998, ApJ, 494, 799

\refer Fullerton A.W., Massa D.L., Prinja R.K., 2006, ApJ, 637, 1025

\refer Grunhut J.H., Wade G.A., Marcolino W.L.F., Petit V., Henrichs H.F., Cohen D.H., Alecian E., Bohlender D., Bouret J.-C., et al., 2009, MNRAS, 400, L94

\refer Hanson M.M., Kudritzki R.-P., Kenworthy M.A., Puls J., Tokunaga A.T., 2005, ApJS, 161, 154

\refer Heap S.R., Lanz T., Hubeny I., 2006, ApJ, 638, 409

\refer Herrero A., Kudritzki R.P., Vilchez J.M., Kunze D., Butler K., Haser S., 1992, A\&A, 261, 209

\refer Hillier D.J, 1991, A\&A, 247, 455

\refer Howarth I.D., Siebert K.W., Hussain G.A.J., Prinja R.K., 1997, MNRAS, 284, 265

\refer Howarth I.D., Walborn N.R., Lennon D.J., Puls J., 2007, MNRAS, 381, 433

\refer Lamers H.J.G.L.M., Snow T.P., Lindholm D.M., 1995, ApJ, 455, 269

\refer L\'epine S., Moffat A.F.J., St-Louis N., Marchenko S.V., Dalton M.J., Crowther, P.A., Smith L.J., Willis A.J., Antokhin I.I., Tovmassian G.H., 2000, AJ, 120, 3201

\refer Marcolino, W.L.F., Bouret J.-C., Martins F., Hillier D.J., Lanz T., Escolano C., A\&A, 498, 837

\refer Markova N., Puls J., 2008, A\&A, 478, 823

\refer Martins F., Schaerer D., Hillier D.J., 2002, A\&A, 382, 999

\refer Martins F., Schaerer D., Hillier D.J., Heydari-Malayeri M., 2004, A\&A, 420, 1087

\refer Martins F., Hillier D.J., Paumard T., Eisenhauer F., Ott T., Genzel R., 2008, A\&A, 478, 219

\refer Martins F., Hillier D.J., Bouret J.C., Depagne E., Foellmi C., Marchenko S., Moffat A.F., 2009, A\&A, 495, 257

\refer Martins F., Donati J.-F., Marcolino W.L.F., Bouret J.-C., Wade G.A., Escolano C., Howarth I.D., 2010, MNRAS, 407, 1423

\refer Najarro F., Krabbe A., Genzel R., Lutz D., Kudritzki R. P., Hillier D.J., 1997, A\&A, 325, 700

\refer Najarro F., Hillier D. J., Puls J., Lanz T., Martins F., 2006, A\&A, 456, 659

\refer Najarro F., Figer D.F., Hillier D.J., Geballe T.R., Kudritzki R.P., 2009, ApJ, 691, 1816

\refer Penny L.R., 1996, ApJ, 463, 737

\refer Prinja R.K., Barlow M.J., Howarth I.D., 1990, ApJ, 363, 607

\refer Puls J., Kudritzki R.-P., Herrero A., Pauldrach A.W.A., Haser S.M., Lennon, D.J., Gabler R., Voels S.A., Vilchez J.M., Wachter S., Feldmeier A., 1996, A\&A, 305, 171

\refer Puls J., Markova N., Scuderi S., Stanghellini C., Taranova O.G., Burnley A.W., Howarth I.D., 2006, A\&A, 454, 625

\refer Puls J., Vink J.S., Najarro, F., 2008, A\&ARv, 16, 209 

\refer Repolust T., Puls J., Hanson M.M., Kudritzki R.-P., Mokiem M.R., 2005, A\&A, 440, 261

\refer Ryans R.S.I., Dufton P.L., Rolleston W.R.J., Lennon D.J., Keenan F.P., Smoker J.V., Lambert D.L., 2007, MNRAS, 336, 577

\refer Sim\'on-D\'iaz S., Herrero A., Esteban C., Najarro F., 2006, A\&A, 448, 351

\refer Sim\'on-D\'iaz S., Herrero A., 2007, A\&A, 468, 1063

\refer Sim\'on-D\'iaz S., Herrero A., Uytterhoeven K., Castro N., Aerts C., Puls J., 2010, APJ, 720, L174

\refer Sundqvist J.O., Puls J., Feldmeier A., 2010, A\&A, 510, A11

\refer Slettebak A., Collins G.W.II, Parkinson T.D., Boyce P.B., White N.M., 1975, ApJS, 29, 137

\refer Trundle C., Lennon D.J., Puls J., Dufton P.L., 2004, A\&A, 417, 217

\endrefer           
\end{document}